\documentclass[
aip, 
jmp,
amsmath, amssymb, 
reprint,
]{article}

\usepackage{graphicx}
\usepackage{makeidx}
\usepackage{multirow}
\usepackage{amssymb}
\usepackage{amsfonts}
\usepackage{amsmath}
\usepackage{amsthm}
\usepackage{natbib}

\usepackage{graphicx}
\usepackage{dcolumn}
\usepackage{bm}

\usepackage[utf8]{inputenc}
\usepackage[T1]{fontenc}
\usepackage{mathptmx}
\usepackage{amsfonts}
\usepackage{bbm}
\usepackage[bottom]{footmisc}

\begin{document}

\title{A simple quantum model linked to a theory of decisions}

\author{Inge S. Helland
\\ Department of Mathematics, University of Oslo \\ P.O.Box 1053 Blindern, N-0316 Oslo, Norway
\\ Orcid: 0000-0002-7136-873X\\ Telephone: (047)93688918\\ E-mail: ingeh@math.uio.no}

\date{\today}

\maketitle

\begin{abstract}
This article may be seen as a summary and a final discussion of the work that the author has done in recent years on the foundation of quantum theory. It is shown that quantum mechanics as a model follows under certain specific conditions from a quite different, much simpler model. This model is connected to the mind of an observer, or to the joint minds of a group of communicating observers. The model is based upon conceptual variables, and an important aspect is that an observer (a group of observers) must decide on which variable to measure. The model is then linked more generally to a theory of decisions. The results are discussed from several angles. In particular, macroscopic consequences are treated briefly.

\end{abstract}

\underline{Keywords:} Accessible variables; conceptual variables; decisions; maximally accessible conceptual variables; models; reproducing quantum mechanics; simple model.
\bigskip

\section{Introduction}
\label{Sec1}

Through a book [1] and two recent articles [2, 3], this author has developed a different approach towards quantum mechanics. Instead of taking the abstract Hilbert space apparatus as a point of departure, I start with what is called  conceptual variables and some simple ideas connected to these. This, together with a general epistemic interpretation, gives a completely new view on the foundation of quantum theory. This new foundation can be communicated to outsiders, and not least, it may in some sense be seen to solve the serious problem of `understanding' quantum mechanics. Thus these are really quite radical ideas.

As an author, I have not yet received any comments, neither negative nor positive on this book and the articles. The articles have been published in recognized theoretical physical journals, so at least from this point of view, there must be something to the ideas expressed there. How can then the lack of reactions be explained? Admittedly, the papers require a new way of thinking, and a new conceptual framework to learn.  I choose to rely on something that Carlo Rovelly recently said in a video interview: The difficult thing about learning new things, is to dislearn what you already have learned.

Therefore it will be tried here to explain the essence of the ideas as simply and precisely as possible, basing the whole thing on conceptual variables and the associated decisions that we may make. It is all first formulated in terms of a simple measurement model. The derivation of quantum mechanics from this model under a certain condition is given explicitly, and some consequences of the theory are discussed.

Then an important observation is made: The mathematical theory used in the above derivation, is also valid when measurement variables are replaced by decision variables. Thus a quantum theory of decisions also follows from this development.

The plan of the article is: In Section 2 the basic concepts are explained, and some literature on quantum decisions is briefly mentioned. Then Section 3 contains the main model of this article. The model is described and discussed. In Section 4 it is shown, partly using earlier work of the author, that quantum mechanics under certain conditions is implied by the model. This is considered to be an important result. So-called paradoxes are addressed in Section 5, and Section 6 gives two important cases where this derivation of quantum axioms can be illustrated. Macroscopic decisions are discussed in Section 7, while Section 8 gives the final discussion.

\section{Decisions, conceptual variables, and some corresponding literature}

Of course, the phenomenon of decisions is crucial to all of us. We go through life making decision after decision. Some of our decisions are automatic. Many of the decisions may partly be caused by our upbringing, our past experiences, people that we have met, or people that we look up to; it might be that some of these people have made similar decisions before. Our decisions may also depend in general on the context that we are in, say the social context. But still there will be something important left. In this article, a single actor $B$ in some context is first considered, and it is assumed that his decisions partly depend on the variables that he has in his mind at the moment when he makes these decisions.

These variables will be seen as primitive elements, and they will be called conceptual variables. In a physical context, a conceptual variable $\eta$ is any variable that can be used in the modelling of some system.

In [1] and [2] such variables were used as important elements in a derivation of essential parts of quantum mechanics, using technical assumptions related to symmetry, more concretely, group actions and group representations. The decision involved may simply be the choice of what to measure. The relevant decisions may be made by an actor $B$, or by a group of communicating actors.

But the theory of decisions addressed in this article is more general. As such, it is discussed in more detail in [1], and some of this will be recapitulated below.

In the process of making decisions, we will in general be far from perfect. So also the actor $B$. In [3] the following is shown, using arguments connected to one of Bell's inequalities: Any actor may be limited when making a decision, and this limitation may be due to the fact that he is not, while making the decision, able to keep enough variables in his mind.

Also, when making decisions, $B$ may need to observe his environment, and try to interpret these observations. Then all this is consistent with the Convivial Solipsism proposed by Zwirn [4]: Every description of the world must be related to the mind of some actor. But other persons also have their minds, and different persons can communicate. In a quantum mechanic setting we need observations and variables connected to these observations. That is a measurement theory. This will be explicitly discussed below.

Before making decisions, it may be important to achieve as much knowledge as possible. The process of achieving knowledge - an epistemic process - is the basis for the book [1].

Alternative to a single actor making decisions, we can look at joint decisions made by a communicating group of people. The same theory can be used, with the following limitation: All variables must, to enable communication, be possible to define in terms of words.

Note that in any case, the process of making decisions will be assumed to be intimately connected to the conceptual variables that the actor or actors have in their minds at the moment of decision.

A conceptual variable can be a scalar or a vector, and also more general.

The purpose of this article is to describe in simple terms a model for the process of making decisions, and from this a rather elementary quantum-like model, a model which can be described without any advanced mathematics. The point is now that this model can be seen as extending ordinary quantum mechanics, in the sense that the ordinary quantum model can be derived under the model by making some additional assumptions [1, 2].

There is a large literature on quantum decisions, and it is all related to what I will say below. I see no reason for going into details here. A pioneering article was the one by Yukalov and Sornette [5], a couple of authors who have since written many articles on the subject. The area is surveyed and tied to the literature on quantum models of cognition in [6]. The latter literature is  reviewed recently  in [7]. Weak points of the models and open questions are also discussed in these articles.

This literature can be seen as a background and motivation for my development, but it is in no way assumed in formulating the model below.

\section{A simple measurement or decision model}

In a physical context, the variable $\theta$ can be any physical variable like time, space variable, momentum or spin component. A crucial assumption is that $\theta$, in addition to its physical existence, before, during and after measurement also exists in the mind of an actor $B$, or in the joint minds of some communicating actors. The variable $\theta$ may be assumed to vary over some space $\Omega_\theta$.

A completely different situation, also covered below, is that $\theta$ is a decision variable. Assume that $B$ is to choose between the prospects $\{\pi_k ; k=1,...,r\}$ (a larger set of prospects may also be considered). Define $\theta=k$ as corresponding to the case where $\pi_k$ is chosen. A `measurement' is then thought of as the process of realizing the decision. 

In any case, a set of possible conceptual variables $\{\theta^a ; a\in \mathcal{A}\}$, where $\mathcal{A}$ is some index set, may exist in the mind of the actor.

The only property that these variables are required to satisfy, is: If $\theta$ is a conceptual variable, and $f$ is some function on $\Omega_\theta$, then $\xi=f(\theta)$ is also a conceptual variable.

A conceptual variable is called \emph{accessible} if it is possible to find some accurate value of $\theta$ at some point of time. An example of a physically inaccessible variable may be the two-dimensional vector (position, momentum), another may be the full spin vector of a particle. In the simple model for decisions mentioned above, a decision variable is called accessible if it really is possible to carry out the decision connected to this variable.

A partial ordering among the conceptual variables, and also among the accessible ones, may be defined: Say that the variable $\theta$ is less than the variable $\lambda$ if $\theta=f(\lambda)$ for some non-invertible function $f$. I will assume that, if $\lambda$ is accessible and $\theta$ is less than $\lambda$, then $\theta$ is accessible.

An essential part of the model is now: Assume that there in some given context exists an inaccessible variable $\phi$ such that all accessible variables can be seen as functions of $\phi$. We can write $\theta=f(\phi)$, where $f=f_\theta$, or simply $\theta=\theta (\phi)$.

In a physical context, the variable $\phi$ may be rather concrete, say a particle's spin vector. The psychological case is  more complicated.

In this case one may assume that $\phi$ in contained in the subconsciousness of $B$. The actor $B$ himself will not be able to know $\phi$ consciously, but a friend of $B$, knowing some psychology, may guess parts of $\phi$.

Using a religious language, one may say that $\phi$ is known by God. It must be stressed, however, that what will be said in this article, does not in any way depend on religion. (Although concrete ideas on religion [8] definitely have inspired the writings of this author.)

Assuming the existence of such a $\phi$ may in general be seen as a simplification. But it is a useful simplification, and it determines the model.

The first thing that then will be done, is to introduce the important notion of \emph{maximally accessible} conceptual variables. In the partial ordering defined above, all accessible conceptual variables are dominated by $\phi$. Therefore maximal such variables exist by assuming Zorn's lemma. In a simple physical setting, it follows from Heisenberg's inequality that variables like position, momentum - and spin components - are maximal.

Then it will be defined what is meant by \emph{related} maximal variables, say $\theta$ and $\eta$. These are said to be related if there exists a transformation $k$ in $\Omega_\phi$ such that $\eta(\phi)=\theta(k\phi)$. If no such $\phi$ and $k$ may be found, $\theta$ and $\eta$ are said to be essentially different.

This is all there is to the model. Note that the model may be associated with a concrete context, physically or otherwise.

In agreement to what has been said before, a decision variable will be inaccessible if it is impossible for the actor to make the corresponding decision. The model again assumes a large inaccessible variable $\phi$ such that all accesible ones are functions of $\phi$. From this, also related maximally accessible variables may be defined.

In general, a symmetry version of the model may be defined by assuming that group actions, transitive and with a trivial isotropy group are defined on the maximally accessible conceptual variables, and that this group induces an invariant measure.

For a variable taking a finite number of values, we can use the cyclic group.

In Section 10 of [3] a general theory of a mind's limitation is described. This theory is particularly relevant during the process of making decisions. Theorem 2 in [3] says the following: Assume that the individual $B$ has two related maximally accessible variables $\theta$ and $\eta$ in his mind. Assume that the transformation $k$ of $\Omega_{\phi}$ belongs to a class $K$ which has a certain property relative to the function $\theta(\cdot)$ (permissibility; see below). In this situation $B$ can not have in mind any other maximally accessible variable which is related to $\theta$, but essentially different from $\eta$. This result is shown in [3] to be a consequence of what is called Theorem 1 and Theorem 2 in the next section.

\section{Derivation of quantum mechanics}

Assume now, in a concrete context, a symmetry version of the model above. The model relies on maximally conceptual variables connected to an observer or to a group of communicating observers. The maximality notion implies that the observer(s) only can measure one of these variables. He/ they has/ have to make a choice, a decision. Thus the model is tied to decisions. An important mathematical result is that essential elements of the quantum formalism follow if the observer(s) in his/ their mind has/ have two related maximally accessible variables, and some specific further symmetry conditions are imposed. This is the content of Theorem 1 and Theorem 2 below.

But after this, the link to decisions is strengthened. The conceptual variables in Theorem 1, originally thought of as physical, measurable variables, can also be replaced by decision variables. The same mathematical theory works in the two cases. Thus a quantum theory of decisions is also implied by these mathematical results.

In the following, I will use 3 Sections to discuss the mathematical theory that leads to quantum mechanics from the perspective of the simple model; then I will return to decisions in Section 7.

In [2] this theorem is proved:
\bigskip

\textbf{Theorem 1.} \textit{Let $\theta$ and $\eta$ be two related maximally accessible conceptual variables, and let $G$ be the group acting on $\theta$. Make the following assumption:}

\textit{There exists a unitary irreducible representation $U(\cdot)$ of $G$ such that the coherent states $U(g)|\theta_0\rangle$ are in one-to-one correspondence with the values of $g\in G$ and hence with the values of $\theta$.}

\textit{Then there exists a Hilbert space $\mathcal{H}$ connected to the situation, and to every accessible conceptual variable there can be associated a symmetric operator on $\mathcal{H}$.}
\bigskip

A symmetric operator $A$ is one for which $\langle \alpha, A\beta\rangle=\langle A\alpha,\beta\rangle$ for all $|\alpha\rangle$, $|\beta\rangle$ in the domain of $A$. There is a close connection between the property of being symmetric and the property of being self-adjoint [9].
\bigskip

\underline{Important note.}

In the setting behind this theorem, one must explicitly exclude the trivial case where $\phi=(\theta,\eta)$, and $k$ just exchanges these two values. If this was allowed, all maximal variables would have been related. The point is that the proof of Lemma 1 in [2] cannot be carried out for this case.
\bigskip

In Theorem 1, $\theta$ and $\eta$ can be any variables. In this Section it is important that they are physical variables. But they can also be decision variables, which gives a new version of the theory. 

It is essential that the two variables are related. The important content of Theorem 1 is not that the Hilbert space exists, but that every conceptual variable has an operator associated with it. In [1], these operators are first constructed for the maximal variables, for other variables the construction uses the spectral theorem for these operators.

The task now is to derive the rest of quantum mechanics. Then we first need a new definition.
\bigskip

\textbf{Definition 1.} \textit{The function $\theta(\cdot)$ on the space $\Omega_\phi$ upon which a group $K$ of transformations $K$ is defined is said to be \emph{permissible} if the following holds: $\theta(\phi_1)=\theta(\phi_2)$ implies $\theta(k\phi_1)=\theta(k\phi_2)$ for all $k\in K$.}
\bigskip

This notion is studied thoroughly in [10]. The main conclusion is that if $\theta(\cdot)$ is permissible, then there is a group $G$ acting on the image space $\Omega_\theta$ such that $g(\theta(\phi))$ is defined as $\theta(k\phi)$; $k\in K$. The mapping from $K$ to $G$ is an homomorphism. If $K$ is transitive on $\Omega_\phi$, then $G$ is transitive on $\Omega_\theta$. (Lemma 4.3 in [1].) 

A first possible application of the concept of permissibility, is the following: Can the simple model of Section 3 really reproduce quantization? This can perhaps be approached by taking a group $K$ acting on $\Omega_\phi$ as a point of departure. One can ask whether the mechanism of model reduction can lead to quantization. Specifically, if $\theta(\cdot)$ is not permissible, it may happen that a reduced version of $\theta$ is permissible, and that this reduced version is discrete. This issue was briefly discussed in [1], but requires much further investigation.

The following important result is Theorem A1 in [3]:
\bigskip

\textbf{Theorem 2}
\textit{Assume that the function $\theta(\cdot)$ is permissible with respect to a group $K$ acting on $\Omega_\phi$. Assume that $K$ is transitive and has a trivial isotropy group. Let $T(\cdot)$ be a unitary representation of $K$ such that the coherent states $T(k)|\psi_0\rangle$ are in one-to-one correspondence with $k$. For any transformation $t\in K$ and any such unitary representation $T$ of $K$, the operator $T(t)^\dagger A^\theta T(t)$ is the operator corresponding to $\theta'$ defined by $\theta'(\phi)=\theta(t\phi)$.}

\bigskip

By using this result in the same way as Theorem 4.2 is used in [1], a rich theory follows. The results will first be limited to the case where $\theta$ is a discrete conceptual variable. Then one can show:

1) The eigenvalues of $A^\theta$ coincide with the values of $\theta$.

2) The variable $\theta$ is maximally accessible if and only if the eigenvalues of $A^\theta$ are non-degenerate.

3) For the maximal case the following holds in a given context: a) For a fixed $\theta$ each question `What is the value of $\theta$?' together with a sharp answer `$\theta=u$' can be associated with a normalized eigenvector of the corresponding $A^\theta$. b) If there in the context is a set $\{\theta^a ; a\in \mathcal{A}\}$ of maximally accessible conceptual variables (these must by Theorem 1 be related to each other) one can consider all ket vectors that are normalized eigenvectors of some operator $A^{\theta^a}$. Then each  of these may be associated with a unique question-and-answer as above. 

4) For any accessible variable $\theta$, the eigenspaces of the corresponding operator can be associated with a question-and answer set as above. The eigenvalues are the possible values of $\theta$.
\bigskip

The pure states corresponding to a question-and-answer are abundant and easy to interpret. For instance in the spin component case, it can be proved [1] that these constitute all possible pure states. To characterize in general situations where this is the case, is an open problem, a problem that was presented to the quantum community in [11]. So far, no comments on this important problem have been received.

The mixed states are as usual formed from probability distributions over pure states.

What is left now to construct the full text book version of quantum mechanics, is to derive the Born formula and the Schr\"{o}dinger equation from certain assumptions. For this, the reader is referred to Chapter 5 in [1].

By considering continuous variables as limits of discrete states, a general theory can be constructed also for these variables; see Section 5.3 in [1]. Note that Theorem 1 and Theorem 2 above also are valid for these variables.

\section{Linear combination of states, an entangled state and so-called paradoxes}

I will here give some discussions related to the derivation of the quantum formulation from Theorem 1 and Theorem 2.

A general ket vector $|v\rangle\in\mathcal{H}$ is always an eigenvector of \emph{some} operator associated with a conceptual variable. It is natural to conjecture that this operator at least in some cases can be selected in such a way that the accessible variable is maximally accessible. Then $|v\rangle$ is in a natural way associated with a question-and-answer pair. It is of  interest that H\"{o}hn and Wever [12] recently derived quantum theory for sets of qubits from such question-and-answer pairs; compare also the present derivation. Note that the interpretation implied by such derivations is relevant for both the preparation phase and  the after-measurement phase of a physical system.

One may also consider linear combinations of ket vectors from this point of view. Let the maximally accessible variable $\theta^b$ admits values $u_i^b$ that are single eigenvalues of the operator $A^b$, uniquely determined from $\theta^b$. Let $|b;i\rangle$ be the eigenvector associated with this eigenvalue. Then one can write
\begin{equation}
|b;i\rangle=(\sum_j |a;j\rangle\langle a;j|)|b; i\rangle =\sum_j \langle a;j|b;i\rangle |a;j\rangle .
\label{u0}
\end{equation}

This state can be interpreted in terms of the question `What is the value of $\theta^b$?' with the sharp answer `$\theta^b =u_i^b$'. But if one tries to ask questions about $\theta^a$ for a system where the observer or the set of observers is in this state, the answer is simply `I (we) do not know'.

Also, entangled states may be considered from this point of view. Consider two qubits 1 and 2, each having possible spin values in some fixed direction given by +1 and -1, and look at the entangled state
\begin{equation}
|\psi\rangle =\frac{1}{\sqrt{2}}(|1+\rangle\otimes |2-\rangle - |1-\rangle \otimes |2+\rangle).
\label{ent}
\end{equation}

As discussed in [1] and references given there, this is an eigenvector of the operator for the conceptual variable $\delta =\theta_x\eta_x+\theta_y\eta_y+\theta_z\eta_z$ corresponding to the eigenvalue -3. Here $\mathbf{\theta}=(\theta_x ,\theta_y , \theta_z )$ and $\mathbf{\eta}=(\eta_x ,\eta_y , \eta_z )$ are the total spin vectors of the two qubits, and $\delta$ is accessible to an observer observing both qubits. Note that $\delta =-3$ implies $\theta_x\eta_x=\theta_y\eta_y=\theta_z\eta_z=-1$, which again implies that $\theta_a =-\eta_a$ for any direction $a$. It is trivial that this state can be interpreted in terms of the question `What is $\delta$?' together with the sharp answer `$\delta=-3$'. (The other possible value for $\delta$ is -1, which corresponds to a three-dimensional eigenspace for the operator.) It is also interesting that the observer observing both qubits may be any person with a large enough Hilbert space associated with his mind. This person may make any decision that involves his own accessible variables, but in doing so, he has the limitations discussed in [3].

From a general point of view it may be considered of some value to have an epistemic interpretation which is not necessarily tied to a strict Bayesian view (see for instance [13]  on this). Under an epistemic interpretation, one may also discuss various ``quantum paradoxes'' like Schr\"{o}dinger's cat, Wigner's friend and the two-slit experiment. 

In Example 1 and Example 2 below I limit myself to quantum states that can be interpreted in terms of question-and-answers for maximally accessible variables. In the contexts discussed, this gives an understanding where the relevant observers are not able to distinguish between superposition and mixture. This limitation may be said to imply a concrete new variant of quantum mechanics, a variant relative to which socalled paradoxes may be explained.
\bigskip

\textbf{Example 1.} \textit{Schr\"{o}dinger's cat.} The discussion of this example concerns the state of the cat just
before the sealed box is opened. Is it half dead and half alive?

To an observer outside the box the answer is simply: ``I do not know''. Any accessible e-variable
connected to this observer does not contain any information about the status of life of the cat. But on
the other hand – an imagined observer inside the box, wearing a gas mask, will of course know the
answer. The interpretation of quantum mechanics is epistemic, not ontological, and it is connected to the
observer. Both observers agree on the death status of the cat once the box is opened.
\bigskip

\textbf{Example 2.} \textit{Wigner’s friend.} Was the state of the system only determined when Wigner learned the
result of the experiment, or was it determined at some previous point?

My answer to this is that at each point in time a quantum state is connected to Wigner’s
friend as an observer and another to Wigner, depending on the knowledge that
they have at that time. The superposition given by formal quantum mechanics corresponds to a `do not
know' epistemic state. The states of the two observers agree once Wigner learns the result of the
experiment.
\bigskip

\textbf{Example 3.} \textit{The two-slit experiment.} This is an experiment where all real and imagined observers
communicate at each point of time, so there is always an objective state. 

Look first at the situation when we do not know which slit the particle goes through. This is 
a `do not know' situation. Any statement to the effect that the particles somehow pass through both
slits is meaningless. The interference pattern can be explained by the fact that the particles are (nearly)
in an eigenstate in the component of momentum in the direction perpendicular to the slits in the plane
of the slits. And by de Broglie's theory, momentum is connected to a wave. On the other hand, if an observer finds out which slit the particles go through, the state changes into an eigenstate for the position in that direction. In either case the state is an epistemic state for each of the communicating observers, which might indicate that it in some sense can be seen as an ontological state. From one point of view this can perhaps be seen as a state of the screen and/or the device to observe the particle, not as an ontological state of the particle itself. The state of the `particle', specifically if it can be seen as a wave or as a genuine particle, depends on the choice of measurement made by the observer(s).

\section{Two special cases}

The purpose of this Section is to make my general approach to quantum mechanics concrete in two well known cases.

a) Consider the spin of a spin 1/2 particle, and let $\theta^a$ be the spin component in direction $a$. Then the group $G$ consists just of holding fixed or switching the two possible values $-1$ and $+1$. To construct the Hilbert space and the relevant operators, we need according to Theorem 1 spin components in two directions $a$ and $b$. Let $\phi$ be the unit vector pointing in a certain direction in the plane that is spanned by the vectors $a$ and $b$. Let then $K$ be the rotation in this plane, and let $\theta^a$ be the sign of the component of $\phi$ in the $a$-direction. Then according to Proposition 2 in [2], $\theta^a (\cdot)$ is permissible. To satisfy the reqirements of Theorem 1, we also need to construct a suitable representation of the group $G$. This group contains two elements, the unit $e$ and the group element $g$ switching the values $-1$ and $+1$. The Hilbert space corresponding to $\theta^a$ is just the space spanned by the vectors $(0,1)'$ and $(1,0)'$, and for $|\theta^a\rangle$ being one of these vectors, we define the coherent states
\begin{equation}
U(g)|\theta^a\rangle = e^{-i\theta^a}|g\theta^a\rangle .
\label{Ug}
\end{equation}
This gives two coherent states. If then $|\theta^a\rangle$ is a linear combination of $(0,1)'$ and $(1,0)'$, let $U(g)|\theta^a\rangle$ be the corresponding linear combination of the two coherent states above. From this, $U(g)$ can be found as a $2x2$ unitary matrix. $U(e)$ is just the unity. It should be clear then, that by choosing some initial state $|\theta^a_0\rangle$, the states $U(h)|\theta^a_0\rangle ;h\in G$ are in one-to-one correspondence with $h\in G$.

A further discussion of this example is given in [2].
\bigskip

b) Look at a single particle at some fixed time $t$. Let $\theta$ be its position and $\eta$ its momentum. In [1], Section 5.3, this example is considered as a limit of a discrete case, and the operators in a well-known Hilbert space are constructed. One possibility for the group $K$ is to take the Heisenberg-Weyl group [14]. According to [14], an element $x$ of the corresponding algebra can be written
\begin{equation}
x=x_1 e_1 + x_2 e_2 + s e_3 ,
\label{xx}
\end{equation}
where $e_1, e_2$ and $e_3$ are operators satisfying the commutation relations
\begin{equation}
[e_1 ,e_2]=e_3,\ \ [e_1,e_3] = [e_2, e_3] =0,
\label{xxx}
\end{equation}
and $x_1, x_2$ and $s$ are real numbers. Specifically, $e_1$ is the operator associated with $i(\hbar)^{-1/2}\eta$, $e_2$ is the operator associated with $i(\hbar)^{-1/2}\theta$, and we can define the scalars  $x_1 = -\hbar^{1/2}\theta$ and $x_2 =\hbar^{1/2}\eta$. We can then let $\phi$ be $x$, and let the corresponding group element to be
\begin{equation}
k=\mathrm{exp}(-(\hbar)^{-1/2}x) =\mathrm{exp} (-(\hbar)^{-1/2}se_3)\mathrm{exp}(\theta e_1-\eta e_2).
\label{xx1}
\end{equation}

Then $\theta$ will be a permissible function of $\phi$, and the group element in the $\theta$-space can be taken to be
\begin{equation}
g=\mathrm{exp}(\theta e_1).
\label{xx2}
\end{equation}
Furthermore, it is easy to see that $\theta$ and $\eta$ are related. A change from $e_1$ to $-e_2$ can be obtained by a rotation in the subspace spanned by $e_1$ and $e_2$.

The standard textbook treatment of this example follows.

\section{Macroscopic decisions}

The previous 3 Sections required some more mathematics, and a reference to several recent or previous works. By contrast, the model of Section 3 was relatively simple. And, importantly, the model is relevant both for microscopic and macroscopic situations. I will here discuss aspects of such model considerations in relation to decisions, and give some philosophy around decisions in general.

As said in the introduction; we go through life making decision after decision. At least some of our decisions correspond to concrete choices made specifically by ourselves. These decisions are modelled in Section 3 by assuming that we, as actors, have certain conceptual variables in our minds, and that our decisions are related to these variables, either by taking the variables as decision variables, or, in relation to measurements, choosing which maximal variable to measure.

Decisions may be made by a single person, or by a group of communicating persons. In the last case the persons in the group must exchange information, and information is exchanged by using words.

Let us go back to the model of Section 3 in connection to the decisions made by a group of people. It is no problem to assume that the group has common conceptual variables $\theta, \xi, \lambda,...$. A greater problem is to establish the existence of a dominating variable $\phi$. In a physical context, $\phi$ can correspond to a pair of variables which from physical reasons must be seen as maximally accessible, but often, $\Omega_\phi$ will be greater than the set of these pairs. (The group $K$ must be another than the group  given by exchanging the variables; conf. the note given after Theorem 1 above.) Different maximal variables are called, after Niels Bohr, complementary. In the last chapter of [1], I try to generalize the complementarity concept to other situations than the purely physical situations. This is useful, but it cannot give everything about common decisions.

At last, some words about political decisions, which is a large, important, but difficult field. Ideally such decisions should be preceded by as much knowledge as possible, and ideally again, this knowledge, to the extent that it is not already available, should be found by scientific investigations. In some cases this may be impractical, but on important issues, like the climate issue, it is imperative.

It is interesting that Niels Bohr's concept of complementarity, deeply connected to physics, also can be seen to have applications in politics. In the language of this article, two conceptual variables are complementary if they are different and both maximally accessible. And conceptual variables can be anything.

A large problem is conflicts between nations and group of nations. Often these have a background in complementary \emph{world views}, a problem discussed in [1]. My general opinion, which I share with many, is that these conflicts ought to be solved by negotiations, not by wars or by injustices made by the stronger part. In the light of complementarity or near complementarity, this may often  be extremely difficult, but at least it demands good political leaders that are capable of taking into account different cultures.

\section{Discussion}

First some brief words about cultures in science. Scientific investigations are made by single researchers or by groups of researchers, who are not always perfect in their decisions. (Compare the general conclusion in [2]). It is also a problem that communication may be difficult in situations which requires researchers from different disciplines. Then a common language may be helpful. One of the goals of the present investigation has been to try to develop such a language in the case where one of the disciplines is related to quantum theory.

The interpretation of quantum mechanics that is implied by the discussion here, is a general epistemic one, where QBism  [13] is just a special case, see arguments in [1]. Even so, ontological aspects can be discussed under this umbrella; see [16].

The arguments of this article have been a mixture of mathematics, physics and psychology. Many scientists, also important ones, have said that they did not quite understand quantum mechanics. In my opinion, to really attempt a serious understanding, we need such a mixture of arguments.

Albert Einstein [15] claimed that quantum mechanics was not a complete theory. I prefer to use the word `model' instead of `theory' for quantum mechanics as such. Together with one of its many interpretations it then constitutes a theory. In all my writings I have advocated a general epistemic interpretation of quantum mechanics.

As I see it, the simple model of Section 3, the symmetry version, constitutes an extension of quantum mechanics as a model. The arguments for this are presented in Section 4.

A very important requirement for the final theory should be that it should be reconcilable with general relativity theory. Using the present approach, this can in principle be formulated in simple terms: The purpose of relativity theory has been to find scientific laws that are the same for all observers, whether they move uniformly or in accelerated motions with respect to each other. In my opinion, the goal of the reconciled theory should be the same, only that the term `observer' should be understood in the meaning of the actor or person described in Section 3 above. Some ideas in that direction are under preparation [17].

\section*{References}

[1] Helland, I.S. (2021). \textit{Epistemic Processes. A Basis for Statistics and Quantum Theory.} 2. edition. Springer, Cham, Switzerland.

[2] Helland, I.S, (2022a). On reconstructing parts of quantum theory from two related maximal conceptual variables.  \textit{International Journal of Theoretical Physics} 61, 69.

[3] Helland, I.S. (2022b). The Bell experiment and the limitation of actors. \textit{Foundations of Physics} 52, 55.

[4] Zwirn, H. (2016). The measurement problem: Decoherence and convivial solipsism. \textit{Foundations of Physics} 46, 635-667. 

[5] Yukalov, V.I. and Sornette, D. (2014). How brains make decisions. \textit{Springer Procedings in Physics} 150, 37-53.

[6] Ashtiani, M. and Azgomi, M.A. (2015). A survey of quantum-like approaches to decision making and cognition. \textit{Mathematical Social Sciences} 75, 49-80.

[7] Pothos, E.M. and Busemeyer, J.R. (2022). Quantum Cognition. \textit{Annual Review of Psychology} 73, 749-778.

[8] Helland, I.S. (2022). On religious faith, Christianity, and the foundation of quantum mechanics. \textit{Eur. J. Theor. Philos.} 2 (1), 10-17.

[9] Hall, B.C. (2013). \textit{Quantum Theory for Mathematicians} Springer, New York.

[10] Helland, I.S. (2010). \textit{Steps Towards a Unified Basis for Scientific Models and Methods.} World Scientific, Singapore.

[11] Helland, I.S. (2019). When is a set of questions to nature together with sharp answers to those questions in one-to-one correspondence with a set of quantum states? arXiv: 1909.08834 [quant-ph].

[12] H\"{o}hn, P.A. and Wever, C.S.P. (2017). Quantum theory from questions.  \textit{Physical Review} A \textbf{95}, 012102.

[13] Fuchs, C.A. (2010). QBism, the perimeter of quantum Bayesianism. arXiv: 1003.5209 [quant-ph].

[14] Perelomov, A. (1986) \textit{Generalized Coherent States and Their Applications} Springer-Verlag, Berlin.

[15] Einstein, A., Podolsky, B. and Rosen, N. (1935). Can quantum-mechanical description of physical reality be considered complete? \textit{Physical Review} 47, 777-780.

[16] Helland, I.S. (2021). Epistemological and ontological aspects of quantum theory. arXiv: 2112.10484 [quant.ph].

[17] Helland, I.S. (2022). Links between relativity theory and a version of quantum theory based on conceptual
variables. In preparation.
\bigskip

\section*{Appendix}

To further illustrate the mixture of mathematics and psychology advocated in this paper, consider the following discussion. The conclusion must perhaps be taken `with a tongue-in-cheek'.

The saying that one cannot have more than two thoughts in one's mind at the same time, is made precise using concepts from the author's published work. This precise version is proved mathematically under certain assumptions.

This Appendix is written in the spirit of Section 3 above, but it is also written in such a way that it can be read independently, without reading Section 3 first.

\subsection*{A1. Introduction}

In some way or other, we all have some kind of mental model behind what we do and what we say. And in what we say, we also must have in mind the persons that we are talking to.

This Appendix is primarily intended for mathematicians. Its purpose is to try to formulate, and prove under some assumptions a precise version of the common saying:
\begin{equation}
\textit{`One cannot have more than two thoughts in one's mind at the same time.'}
\label{1}
\end{equation}

It is then crucial that in the above sentence, `thought' means `maximal thought', a concept that can be made precise. Two different situations seem to contradict this sentence. Either the thoughts are closely related, and then the may in some sense seen as one thought. Or all the thoughts are essentially different. These notions may be made precise, and one thing that I will show, is that these two situations are the only possible situations. Any in between case, say two are related and two others are essentially different, is logically impossible, using some weak assumptions and a model describing the thoughts.

The arguments follow some articles and books that I have publised on mathematical models and on the foundation of scientific theories, among those statistics and quantum theory; see the references above. The resulting theorem can in fact also be applied to any ideas connected to the models behind these theories: Either all are related, or they are all essentially different. In the book [1] I have argued for a relationship between certain ideas.

\subsection*{A2. Thoughts}

First, we will need some sort of idea of what we will mean by a `thought'. It will be taken here as a primitive concept. In practice, it can be anything, from the name of the person that we are confronting now, via some sentence that we have formulated, to a substantial scientific theory.

The only condition is that a thought $\theta$ shall belong to some space of possible thoughts $\Omega_\theta$, and that functions between such spaces can be defined. We also need a concept of `accessible' thoughts, thoughts that can be `realised' in some future. This can be made precise in different ways. One example may be that the thoughts are decision variables, that these decisions are related to something concrete, and that that each decision will have specific impacts on this `something' later. The only formal rquirement that, if $\lambda$ is accessible, and there is a function $f$ from $\Omega_\lambda$ to a new space $\Omega_\theta$, then $\theta = f(\lambda)$ is also accessible.

\subsection*{A3. Preliminaries}

Every theory is based on some assumptions, so also in this case. First, I will assume that Zorn's lemma is valid. (This is related to the axiom of choice.) Next, I will base my reasoning on some specific model of our way of thinking. As any model, it can be critizised, but I will try to argue that it is reasonable, at least in some cases. 

The model runs as follows: A person has, at some fixed moment of time, a number of thoughts in his mind. I will assume that there exists some inaccessible variable $\phi$, varying in a space $\Omega_\phi$, with the following property: For each possible thought $\theta$ there is a function $f$ such that $\theta=f(\phi)$. 

Here is a physical example: An experimentalist will be confronted with a concrete particle at some fixed time $t$, and thinks of measuring either position or momentum of that particle at $t$. By Heisenberg's inequality, the future measurement of the vector (position, momentum) will necessarily be connected to an inaccessible thought $\phi$ in the mind of the experimentalist. The thoughts $\theta$ connected to position and $\xi$ connected to momentum can be seen as functions of $\phi$, and in some sense that will be made precise below, these may be seen as maximal thoughts in this context.

In a more general psychological situation, $\phi$ may be some abstract thought belonging to the subconsciousness of a concrete person $A$. As a model, all thoughts in the mind of $A$ in some context may be thought of as functions of $\phi$. The thought $\phi$ will of course be inaccessible to $A$ himself, but some friend of $A$, knowing some psychology, may be able to guess parts of $\phi$.

In a religious setting, $\phi$ may be thought of as being related to $A$'s faith in God. This is of course just an example; the theory of this paper is not in any way based on religion. 

The model above may, as any model, be seen as a simplification. But I will claim that it is a useful simplification.

Two thoughts $\theta$ and $\xi$ are seen as related if there is an inaccessible $\phi$ and a transformation $k$ on $\Omega_\phi$ such that $\theta = f(\phi)$ and $\xi = f(k\phi)$ for some function $f$. This requires that $\Omega_\theta$ and $\Omega_\xi$ are in one-to-one correspondence. If no such $k$ can be found, we say that $\theta$ and $\xi$ are essentially different. If the choice of variable $\phi$ is fixed, we will say that $\theta$ and $\xi$ are related relative to $\phi$.

Now appeal to Zorn's lemma. We can define a partial ordering of the thoughts of a person $A$: We say that the thought $\xi$ is `less than' the thought $\theta$ if $\xi = h(\theta)$ for some function $h$. This is a partial ordering both among all thoughts in $G$ and also among the accessible thoughts in $G$. Then, since all thoughts in $G$ are dominated by $\phi$, it follows from Zorn's lemma that there exist maximally accessible thoughts. These are important.

\subsection*{A4. Permissibility}

It may be useful to assume that transformations can be defined on a space of thoughts.
\bigskip

\textbf{Definition} \textit{The function $f(\cdot )$ from a space $\Omega_\phi$ upon which a group of transformations $K$ is defined, to a space $\Omega_\theta$, is said to be permissible if the following holds: $f(\phi_1)=f(\phi_2)$ implies $f(k\phi_1) =f(k\phi_2)$ for all $k\in K$.}
\bigskip

It is relatively easy to show that if $f$ is permissible, then group-actions $G$ may be defined on $\Omega_\theta$ by $g(f(\phi))=f(k\phi)$ for $k\in K$. The mapping from K to G here is a homomorphism.

In [3] the following is proved:
\bigskip

\textbf{Theorem A1} \textit{Assume that the individual $A$ has two related maximally accessible variables $\theta$ and $\xi$ in his mind. Then by definition $\xi(\phi) = \theta (k\phi)$ for a transformation $k$ of $\Omega_\phi$.} 

\textit{Assume that a group $K$ of transformations of $\Omega_\phi$ can be defined such that $k\in K$ and $\theta(\cdot)$ is permissible with respect to $K$. In this situation $A$ can not simultaneously have in mind any other maximally accessible variable which is related to $\theta$, but essentially different from $\xi$.}

\subsection*{A5. The main theorem and its proof}

I am now able to formulate and prove my main theorem, based on the assumptions above. Its corollary is intended to be a precise version of the sentence which started this article.

This will require that thoughts are seen as special cases of what I have called conceptual variables. As I have said before, a conceptual variable may be anything, also something as general as a thought. The actions of mathematical groups on these thoughts may of course be discussed, but if one assumes this, the results are quite interesting.
\bigskip

\textbf{Theorem A2} \textit{Assume that a person $A$ in some context at some fixed time has a set of different maximally accessible thoughts in his mind, more than two.}

\textit{Assume that there is a one-to-one correspondence between the spaces behind each of these thoughts.  Then either all the thoughts are related, or they are all essentially different.}
\bigskip

\underline{Proof.} 
Consider  the three thoughts $\theta$, $\xi$, and $\lambda$. Assume that $\theta$ and $\xi$, say, are essentially different, but that $\lambda$ and $\theta$ are related. The only other possibility, except for a symmetric assumption, is that all three are related, or they are all essentially different, I will prove that  the assumption above, following Theorem A1, leads to a contradiction.The proof will be complete if we can show that the transformation $k$ relating $\lambda$ and $\theta$ is contained in a group $K$ such that $\theta(\cdot)$ is permissible with respect to this group.

Let $K$ be a group on $\Omega_\phi$ generated by permuting the elements of $\Omega_\theta$, and keeping any information outside the vector  $(\theta, \lambda, \xi)$   constant, a subgroup of the full permutaion group doing this, such that this subgroup is transitive and has a trivial isotropy group. Since there is a one-to-one correspondence between the two spaces, each permutation of the values of $\theta$ generates a permutation of the values of $\lambda$. By assumption, $\lambda (\phi)=\theta (k\phi)$, and then the transformation $k$ may be taken as one of these permutations.

It is left to prove that $\theta(\cdot)$ is permissible. It is straightforward to prove that the full permutation group is permissible. Assume that $\theta(\phi_1)=\theta(\phi_2)$ for two elements $\phi_1$ and $\phi_2$ of $\Omega_\phi$. Let $t$ be any transformation in $K$. I have to show that $\theta(t\phi_1)=\theta(t\phi_2)$. But this follows, since $t$ by definition permutes the elements of $\Omega_\theta$. If $\phi_1$ and $\phi_2$ belong to the same orbit of $K$, then $t\phi_1$ and $t\phi_2$ also must do that. When $\theta(\cdot)$ is permissible with respect to a group, it is also permissible with respect to any subgroup.

Note that, if one has more than two thoughts in one's mind, any three of them may be the thoughts $\theta$, $\xi$ and $\lambda$.
\bigskip

\textbf{Theorem A3} \textit{Given some context, and given some very technical conditions formulated in the proof below, a person $A$ cannot have more than two essentially different maximally accessible thoughts in his mind at the same time.}
\bigskip

\underline{Proof}
Assume that he has three such thoughts in his mind, call them $\theta$, $\xi$, and $\lambda$. Without loss of generality that there is a group $G_1$ acting upon $\theta$, a group $G_2$ acting upon $\xi$ and a group $G_3$ acting upon $\lambda$, and that these groups are transitive with a trivial isotropy group. Assume also that these groups have representations $U_i (\cdot)$ such that the coherent states $U_i(g_i)|\psi_i\rangle$ that can be constructed are in one-to-one correspondence with $g_i$, $(i=1,2,3)$. The thoughts are essentially different according to an inaccessible variable $\phi$. Define a new inaccessible variable $\phi'=(\phi,\theta, \xi, \lambda)$. With respect to $\phi'$ each pair of thoughts are related. Then according to Theorem 1 in the main text, one can construct three Hilbert spaces, $\mathcal{H}_1$ associated with $\theta$ and $\xi$,  $\mathcal{H}_2$ associated with $\theta$ and $\lambda$, and  $\mathcal{H}_3$ associated with $\lambda$ and $\xi$. And each thought has a a symmetric operator in the relevant Hilbert space attached to it.

Define now $K=G_1\otimes G_2\otimes G_3 \otimes P$, where $P$ is the permutation group between $\theta, \xi$ and $\lambda$. Assume that the technical conditions of Theorem 2 hold, let $t_1$ be the permutation exchanging $\theta$ and $\xi$, $t_2$ be the permutation exchanging $\theta$ and $\lambda$, and $t_3$ be the permutation exchanging $\lambda$ and $\xi$. Finally, let $A_1$ be the operator associated with $\theta$ in $\mathcal{H}_1$, $A_1$ be the operator associated with $\lambda$ in $\mathcal{H}_2$, and $A_3$ be the operator associated with $\xi$ in $\mathcal{H}_3$. Then, according to Theorem 2, there is a unitary similarity transformation connecting $A_1$ and $A_2$,  there is a unitary similarity transformation connecting $A_1$ and $A_3$, and  there is a unitary similarity transformation connecting $A_2$ and $A_3$. But this results in two different formulae for $A_1$, which gives a contradiction.
\bigskip

\underline{Note}
The human mind is very flexible. Taking time into account, we can change thoughts very quickly, and in this way handle many essentially different thoughts. But at each moment of time, we can only think of two essentially different maximally accessible concepts.

\end{document}